\documentclass[fleqn,usenatbib]{mnras}

\usepackage{newtxtext,newtxmath}

\usepackage[T1]{fontenc}

\DeclareRobustCommand{\VAN}[3]{#2}
\let\VANthebibliography\thebibliography
\def\thebibliography{\DeclareRobustCommand{\VAN}[3]{##3}\VANthebibliography}

\usepackage{graphicx}	
\usepackage{amsmath}	
\usepackage{comment}
\usepackage{soul}

\title[HI surveys for testing axion dark matter]{Using HI observations of low-mass galaxies to test ultra-light axion dark matter}

\author[Garland, Masters, \& Grin]{
James T. Garland,$^{1}$\thanks{E-mail: james.garland@astro.utoronto.ca}
Karen L. Masters,$^{2}$
Daniel Grin$^{2}$
\\
$^{1}$David A. Dunlap Department of Astronomy \& Astrophysics,
University of Toronto, 50 St. George Street, Toronto, ON M5S 3H4, Canada\\
$^{2}$Departments of Physics and Astronomy, Haverford College, 370 Lancaster Avenue, Haverford, Pennsylvania 19041, USA
}

\date{Accepted XXX. Received YYY; in original form ZZZ}

\pubyear{2024}

\begin{document}
\label{firstpage}
\pagerange{\pageref{firstpage}--\pageref{lastpage}}
\maketitle

\begin{abstract}
We evaluate recent and upcoming low-redshift neutral hydrogen (HI) surveys as a cosmological probe of small scale structure with a goal of determining the survey criteria necessary to test ultra-light axion (ULA) dark matter models. Standard cold dark matter (CDM) models predict a large population of low-mass galactic halos, whereas ULA models demonstrate significant suppression in this small-scale regime, with halo mass cutoffs of $10^{12}\, \mathrm{M}_{\odot}$ to $10^{7}\, \mathrm{M}_{\odot}$ corresponding to ULA masses of $10^{-24}\,$eV to $10^{-20}\,$eV, respectively. We generate random, homogeneously populated mock universes with cosmological parameters adjusted to match CDM and ULA models. We simulate observations of these mock universes with hypothetical analogs of the mass-limited ALFALFA and WALLABY HI surveys and reconstruct the corresponding HI mass function (HIMF). We find that the ALFALFA HIMF can test for the presence of ULA DM with $m_{a}\lesssim 10^{-21.5}~{\rm eV}$, while WALLABY could reach the larger window $m_{a}\lesssim 10^{-20.9}~{\rm eV}$. These constraints are complementary to other probes of ULA dark matter, demonstrating the utility of local Universe HI surveys in testing dark matter models.
\end{abstract}




\section{Introduction}

The prevailing ``cold dark matter plus cosmological constant, $\Lambda$, ($\Lambda$CDM)'' cosmological model yields compelling explanations for a variety of observational measurements, from the late-time accelerated cosmic expansion, to the large-scale clustering of galaxies \citep{VanWaerbeke:2013eya,Alam:2016hwk,2019PhRvD..99l3505A,2019PhRvL.122q1301A,2021arXiv210513549D}, and the acoustic oscillations of the primordial plasma, probed by the cosmic microwave background (CMB) and galaxy surveys \citep{Beutler:2011hx,Crites:2014prc,Louis:2016ahn,2020MNRAS.498.3470W,2020A&A...641A...6P}. Despite its success at large scales, $\Lambda$CDM predictions face many challenges from small scales. The over-abundance of low-mass galactic halos predicted by $\Lambda$CDM compared to those seen in observations is one such challenge, often known as the ``Missing Satellites Problem'' (MSP, see e.g. \citealt{Papastergis2015,2017ARA&A..55..343B} and references therein for a more complete discussion). 

The need to reconcile $\Lambda$CDM predictions with small-scale structure has fostered the exploration of alternative dark-matter (DM) models, such as self-interacting dark matter (SIDM) \citep{2000PhRvL..84.3760S}, warm DM \citep{Maio:2014qwa,Bose:2016irl,Ludlow:2016ifl}, and fuzzy DM \citep{Hu:2000ke}, all of which are motivated by considerations from high-energy particle theory \citep{2017ARA&A..55..343B}. These models suppress the formation of small-scale structure through DM scattering, enhanced DM velocities at early times, or macroscopic wave-like behavior, respectively \citep{2017ARA&A..55..343B}. Thus, dwarf galaxies could constrain self-interacting and warm dark matter scenarios (e.g. \citealt{2024arXiv240110318N}). Baryon-driven processes (e.g. reionization, supernovae feedback, adiabatic contraction) could ultimately solve the MSP \citep{2017ARA&A..55..343B,2023arXiv231103591E,2023arXiv230900039W}, but alternative DM candidates merit serious consideration, especially given the large number of null results from DM direct-detection experiments targeting weakly-interacting massive particle (WIMP) DM \citep{2019JPhG...46j3003S}.

The ultra-light axion (ULA) is one fuzzy-DM candidate that is well motivated by the compactification of higher-index gauge fields and other considerations from string theory \citep{Dine:1981rt,Weinberg:1977ma, Nambu:1989kh,Peccei:1977hh,PhysRevD.28.1243}. For $m_{ a}\gtrsim 10^{-27}~{\rm eV}$, the dependence of ULA energy density is the same as for matter (i.e. density, $\rho \propto (1+z)^{3}$) during the matter and dark-energy dominated eras. If ULAs compose some or all of the DM, they could resolve some discrepancies between the $\Lambda$CDM model and observations \citep{Khlopov:1985jw,
Frieman:1995pm,Arvanitaki:2009fg,Suarez:2011yf,2012PhRvD..86h3535P,Marsh2016,Hui:2016ltb}. ULAs with $10^{-28}~{\rm eV}\lesssim m_{a}\lesssim 10^{-27}~{\rm eV}$ could mitigate the $S_{8}$ tension, a disagreement in the density-field variance predicted from $\Lambda$CDM fits to CMB data and that measured from galaxy-galaxy weak lensing \citep{Lague:2021frh,Rogers:2023ezo}. ULA-type particles (with non-standard potentials) could also help resolve the Hubble tension \citep{2019PhRvL.122v1301P,Smith:2019ihp,2019arXiv190401016A,2019PhRvD.100f3542L,Hill:2020osr,2020arXiv200910740S,Liu2023}. In addition, ULAs show promise in remedying the MSP and other small-scale challenges for $\Lambda$CDM \citep{Marsh2014}. 

ULA simulations and halo-model calculations demonstrate the presence of a cutoff in the halo mass function, predicting the suppression of the formation of low-mass galaxies, and thus improving agreement between theory and observations \citep{Marsh2014,Schive:2014hza,Veltmaat:2016rxo,Schive:2017biq,2017MNRAS.465..941D,Veltmaat:2018dfz}. The galaxy scale wave-like behavior of ULAs can be thought of as a novel DM pressure (see \citealt{2018PhRvD..97b3529D} and references therein), meaning that the competition between self-gravity and the `fuzziness' of ULAs can be understood through the Jeans instability \citep{Hu:2000ke,2012PhRvD..86h3535P}. The ULA mass, $m_{a}$, thus sets a characteristic Jeans mass scale, which in turn leads to a functional relationship between $m_{a}$ and a cutoff mass in the halo mass function \citep{Marsh2016}. Observationally testing for the existence of this small-scale cutoff could thus be a promising avenue for probing ULA DM.

Various observationally motivated constraints on the ULA model already exist. For a thorough review of progress on testing novel dark matter models with astrophysical probes, see \citet{Buckley2018}. Here we summarize some results on ULA constraints. 

 Observations of CMB acoustic peak height ratios are closely in agreement with the $\Lambda$CDM model, and measurements of the galaxy clustering power spectrum do not show a characteristic cutoff from ULAs at small length scales. These data thus already impose a lower limit to the axion mass of $m_{a}\gtrsim 10^{-25}~{\rm eV}$ if ULAs are to be all of the DM \citep{2006PhLB..642..192A,2015PhRvD..91j3512H,2018MNRAS.476.3063H,2018PhRvD..98h3525P,2021arXiv210407802L,2021arXiv211101199D}. Observations of the Lyman-$\alpha$ forest \citep{Armengaud:2017nkf,Irsic:2017ixq,Kobayashi:2017jcf,2019MNRAS.482.3227N,2019MNRAS.484.4273L,2021PhRvL.126g1302R}, stellar scattering considerations, and infall times of globular clusters around MW-dwarves (among others) critically test the ULA mass range $m_{a}\simeq 10^{-25}~{\rm eV} \to 10^{-20}~{\rm eV }$ (see \citealt{Hui:2016ltb,2019ApJ...871...28B,2018arXiv180800464A,2020PDU....2800503D} and references therein). In some cases these observations put lower limits to ULA masses, while in others, there are modest hints that ULAs improve agreement between theory and observations.  
 
 Looking to the new observational window of gravitational waves, it has been suggested that ULAs in the $10^{-16}~{\rm eV}\lesssim m_{a}\lesssim 10^{-11}~{\rm eV}$ mass band could extract gravitational energy from black holes, which would have implications for gravitational-wave detection and black-hole population statistics \citep{Arvanitaki:2009fg}. Pulsar timing arrays (PTAs) are another promising probe for the imprint of ULAs on the cosmic density field \citep{2018PhRvD..98j2002P,Kato:2019bqz,Hwang:2023odi}. Recent PTA evidence for a stochastic gravitational wave background imposes a powerful constraint to ULAs \citep{NANOGrav:2023hvm,2023arXiv230616227A}. Further analyses of these observations could be used to detect or rule out ULAs in the relevant mass range in coming years. 

 One challenge of testing cosmological models using the properties of low-mass halos is that such halos host small galaxies which can only be seen in the nearby Universe, limiting the survey volume and therefore the statistical rigour with which models can be tested with observational data. Spectroscopic HI surveys provide an interesting avenue to test for statistical measurements of the lowest-mass galaxies, because low-mass galaxies tend to be HI rich and the galaxies in an HI survey tend to be of lower masses and surface brightnesses than those in an optical survey \citep{Huang2012}. 

 As with all observational surveys, given a fixed observing time, spectroscopic HI surveys have to balance depth and sky coverage. Deeper surveys by necessity will cover a smaller area of sky, and therefore provide only small volumes in the nearby Universe where they are sensitive to the lowest-mass objects. The large collecting area of the Arecibo radio telescope enabled the wide area Arecibo Legacy Fast Arecibo L-band Feed Array (ALFALFA) survey \citep{Haynes2018} to relatively quickly survey a large sky area, and hence cover a significant volume even for local, low-mass galaxies. ALFALFA detected hundreds of galaxies with HI masses $M_{\rm HI}<10^8\,\mathrm{M}_\odot$, and even a significant number of galaxies with $M_{\rm HI}<10^7\,\mathrm{M}_\odot$ \citep{Cannon2011}. The new Five-hundred-meter Aperture Spherical radio Telescope (FAST) in China, now the largest single dish telescope in the world, also plans HI surveys. The FAST All Sky HI survey (FASHI) had its first data release earlier this year \citep{Zhang2024} and is predicted to detect three times as many HI sources as ALFALFA (mostly due to larger sky coverage). The development of the Square Kilometer Array (SKA) and its precursors will lead to future large area spectroscopic HI surveys \citep[e.g. WALLABY, and SKA Phase 1; ][respectively]{Koribalski2020,SKA2020} which are planned to be significantly more sensitive, providing even larger numbers of HI detections, and to lower mass limits in the local Universe. 
  
  The HI Mass Function \citep[HIMF, e.g.][]{Zwaan2005,Hoppmann2015,Jones2018} is a statistical tool commonly used in spectroscopic HI surveys, which counts the volume number density of HI halos of a given mass. Measurements of the HIMF across a variety of survey and galaxy environments show a flattening of the low-mass slope, revealing fewer low-mass HI halos than may be found in a $\Lambda$CDM universe. Interpretation of this observation is complicated by baryonic physics which might impact the HI-mass-to-halo mass relationship, although the most extreme ULA impacts on the HIMF should still produce a clear signal. 

Another promising avenue for testing the halo mass function using HI observations is via the HI velocity width function \citep{Zwaan2010, Papastergis2011,Moorman2014}. The results in \citet{Papastergis2011} using this probe clearly indicate a discrepancy with CDM models. In principle this observable probes the DM halo mass more directly than the HI mass, as the link between HI width and DM halo mass should be less impacted by scatter caused by galaxy evolution processes (see \citealt{Paranjape2022} for a recent discussion of use of this as a cosmological probe). However, measuring rotation widths requires higher signal-to-noise ratio than HI masses, which may limit use at lower masses.
  
In this paper, we quantify the spectroscopic HI survey criteria necessary to test ULA masses in the range of $m_a=10^{-24}\,$eV to $10^{-20}\,$eV using the HIMF of nearby Universe, low-mass galaxies. We simulate mock spectroscopic HI surveys as they would appear in a variety of ULA cosmologies by adjusting the cosmological parameters of randomly-generated mock universes. The statistics reconstructed from these mock observations allow us to simulate how the existing ALFALFA survey \citep{Haynes2018} and upcoming WALLABY survey \citep{Koribalski2020} would appear in a ULA cosmology, and quantify the expected reduction of low-mass halos detected relative to the predictions from a traditional CDM model. We find that {using the HIMF from} ALFALFA can test for the presence of ULA DM with $m_{a}\lesssim 10^{-21.5}~{\rm eV}$, while WALLABY could reach the larger window $m_{a}\lesssim 10^{-20.9}~{\rm eV}$. 

We review the theory behind ULA as dark matter candidates in \S \ref{sec:theory}, and summarize the methods we use to generate mock HI surveys in \S \ref{sec:methods}. We show and discuss our results, in \S \ref{sec:results}, and provide a summary/discussion in \S \ref{sec:summary}.

\section{Axion Cosmology} \label{sec:theory}

ULAs are an example of a very light scalar field, which could compose the dark matter or dark energy. Very light scalar fields are fairly ubiquitous in string-inspired scenarios \citep{Svrcek:2006yi}, and may even constitute an ``axiverse'' of ULAs with a broad (and roughly logarithmic) distribution of masses \citep{Arvanitaki:2009fg}. The homogeneous equation of motion for ULAs is given by \citet[][and references therein]{Marsh2016} as 
\begin{equation}
    \ddot{\phi}_{0}+3{H}\dot{\phi}_{0}+m_{a}^{2}\phi_{0}=0,
\end{equation} where $\phi_{0}$ is the homogeneous axion field, ${H}\equiv \dot{a}/{a}$ here is the conformal Hubble parameter, and $a$ is the usual cosmological scale factor (not to be confused with $m_a$ for the mass of the axion).
At early times, the ULA energy density,
\begin{equation}
    \rho_{a}=\frac{m^{2}\phi_{0}^{2}}{2}+\frac{\dot{\phi}_{0}^{2}}{2}
\end{equation} is roughly constant. Subsequently (when ${H}\ll m_{a}/3$), $\phi_{0}$ begins to coherently oscillate, with the cycle averaged density scaling as $\langle \rho_{a}\rangle \propto a^{-3}$ \citep{PhysRevD.28.1243}. This has the same scaling with $a$ as matter. For $m_{a}\gtrsim 10^{-27}~{\rm eV}$, the transition between these two regimes occurs after matter-radiation equality, in which case ULAs behave as matter for the bulk of cosmic structure formation and thus contribute to the DM density of the universe.

Due to the macroscopic scale of the ULA de Broglie wavelength, $\lambda_{a}=1/(m_{a}v_{a})$, where $v_a$ denotes the axion peculiar velocity with respect to the Hubble flow, ULA linear density fluctuations grow at a suppressed rate compared with CDM inhomogeneities (causing novel interference and vortex phenomena on non-linear scales as discussed in \citealt{Schive:2014dra},~\citealt{2017MNRAS.471.4559M}, \citealt{2020MNRAS.494.2027M},~and \citealt{2021ARA&A..59..247H}). Averaging over oscillations in the background (homogeneous) field, $\phi_{0}$, and its perturbations $\delta \phi$, ULAs can be modeled as a fluid with a scale-dependent sound speed given by 
\citep{Hwang:2009js,2020PhRvD.101b3501C,2022arXiv220110238P}
\begin{equation}
    c_{s}^{2}\simeq \frac{k^{2}/(4m_{a}^{2}a^{2})}{1+k^{2}/(4m_{a}^2a^2)}.
\end{equation} On small scales, with wavenumber $k\gg m_{a}a$, we thus have $c_{s}\to 1$, while on large scales, $c_{s}\to 0$, recovering CDM-like behavior. Fractional ULA density fluctuations $\delta_{a}$ thus obey the standard $2^{\rm nd}$-order equation for Jeans instability \citep{Hu:2000ke,2012PhRvD..86h3535P,Marsh2016} like
\begin{equation}
    \ddot{\delta}_{a}+2H\dot{\delta}_{a}+\left[\frac{k^{2}c_{s}^{2}}{a^{2}}-4\pi G \rho_{a}\right]\delta_{a}=0,
\end{equation}with acoustic pressure support preventing perturbation growth for wavenumbers $k>k_{\rm J}$ beyond the comoving ULA Jeans scale, where
\begin{equation}
    k_J = (16\pi G a \rho_{a,0})^{1/4} m_a^{1/2}.
\end{equation} and $\rho_{a,0}$ is the present-day ULA mass density.

This modification to CDM linear growth strongly suppresses the abundance of DM halos with
\begin{equation}
    M\lesssim M_J(z) \equiv \frac{4\pi}{3}200r_J^3\rho_0(z) ,
\end{equation} where $r_J = \frac{2\pi}{k_J}$ and $\rho_{0}(z)$ is the mean cosmic density at redshift, $z$. 

Precise first-principles calculations of halo abundances and correlations in ULA models require computationally expensive simulations (as in \citealt{Schive:2014hza,2017MNRAS.471.4559M,Veltmaat:2018dfz,2020MNRAS.494.2027M}). The halo model \citep{2002PhR...372....1C} offers a sufficient tool to probe ULA models \citep{Marsh2014,2016arXiv160505973M}. A variety of complications emerge, such as the choice of a filter shape for the linear density scale \citep{2013MNRAS.433.1573S,2016arXiv160505973M}, the non-Markovian nature of random walks in the ULA density field \citep{2010ApJ...711..907M,2013MNRAS.428.1774B,2017MNRAS.465..941D}, and the scale-dependence of the spherical density collapse threshold $\delta_{c}$ \citep{2016arXiv160505973M}. Nonetheless, it has been found that ULA halo population is well described using a mass-dependent collapse threshold \citep{2016arXiv160505973M} of
\begin{equation}\label{eq:marsh_deltacrit}
    \delta_{\text{crit}}(M, z) = G(M) \frac{\delta_{\text{crit}}^{\text{CDM}}}{D_{\text{CDM}}(z)},
\end{equation}
with
\begin{align}\label{eq:marsh_growth}
    G(M) &= h_{\text{F}}(x) \exp{[a_3 x^{-a_4}]} + [1-h_{\text{F}}(x)] \exp{[a_5 x^{-a_6}]},\\
     h_{\text{F}}(x) &= \frac{1}{2} \{1-\tanh{[M_J(x-a_2)]}\},\\
     x&\equiv M/M_J,
\end{align}
and 
\begin{equation}
 M_J = a_1 \times 10^8 \left( \frac{m_a}{10^{-22}\ \text{eV}} \right)^{-3/2} \left( \frac{\Omega_m h^2}{0.14} \right)^{1/4} h^{-1} \mathrm{M}_{\odot},
\end{equation} with coefficients\\
$\{a_1, a_2, a_3, a_4, a_5, a_6\}~=~\{3.4, 1.0, 1.8, 0.5, 1.7, 0.9\}$. Here $\delta_{\rm crit}^{\rm}~\simeq~1.686$ is the standard CDM spherical collapse threshold and $D_{\rm CDM}(z)$ is the usual CDM growth function \citep{1992ARA&A..30..499C}. 

The resulting halo population (see, e.g. \citealt{2016arXiv160505973M}) is reasonably well described by the Sheth-Tormen (ST) mass function \citep{Sheth1999}, which is given by
\begin{equation}\label{eq:sheth-tormen}
    f(\nu) = A \sqrt{\frac{a}{2\pi}} \nu \left[ (1+a\nu^2)^{-p} \right] \exp{(-a\nu^2/2)},
\end{equation}
where $A=0.322$, $a=0.707$, and $p=0.3$, with 
$\nu~\equiv~\delta_{\text{crit}}/\sigma(M)~=~G(M)\nu_{\text{CDM}}$. 

\begin{figure}
    \includegraphics[width=\columnwidth]{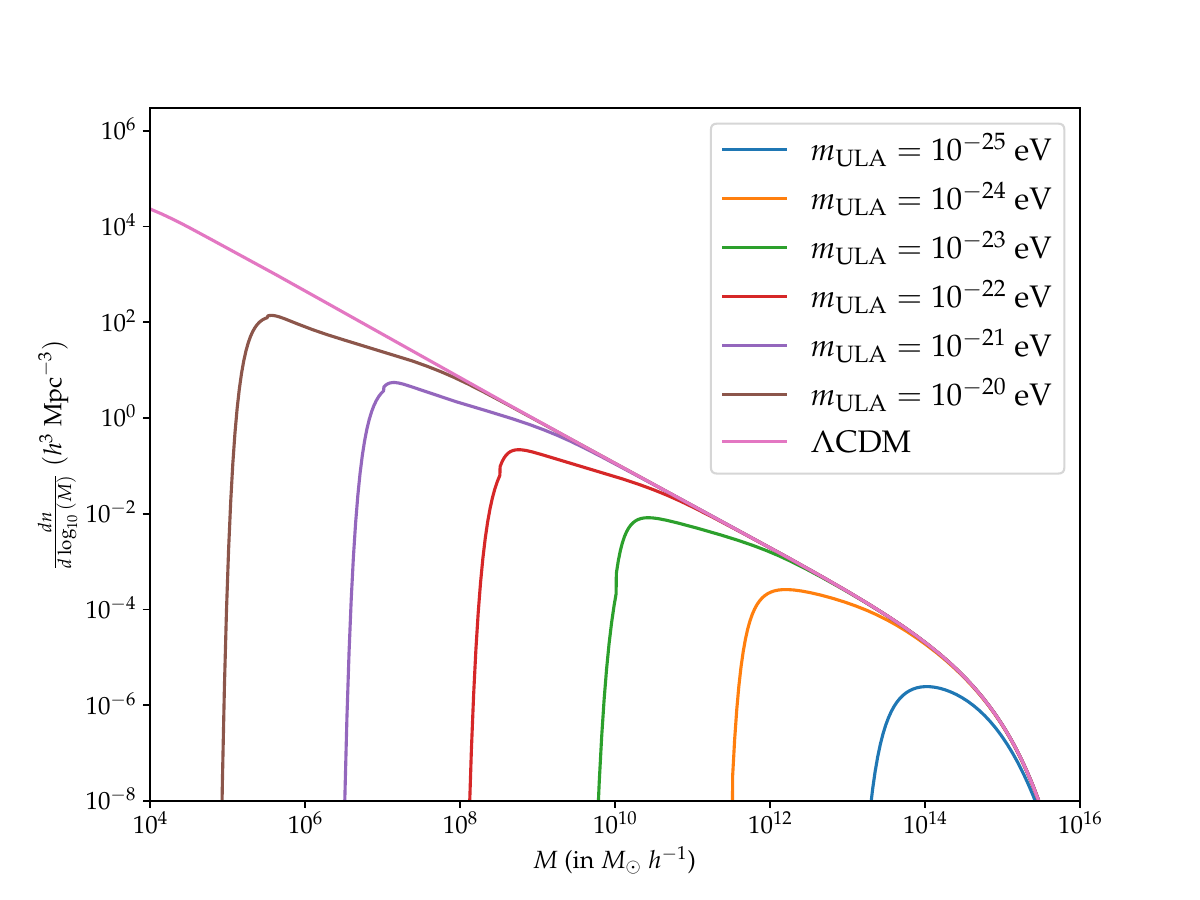}
    \caption{Halo Mass Function (HMF) for $\Lambda$CDM (pink) compared to that found for several ULA axion masses. The subtle discontinuous features appearing just below the turn-over region of our HMFs correspond to distinct features from the analytic fit to the scale-dependent $\delta_{\rm crit}$ used in \citet{2016arXiv160505973M}, and the linear power-spectrum obtained from \textsc{AxionCAMB}.}
    \label{fig:duCutoffs}
\end{figure}

Use of the ST mass function assumes that the threshold for a linear overdensity at the time of halo collapse is given by the scale-independent barrier. More accurate numerical solutions for the mass function were found in \citet{2017MNRAS.465..941D}. We use the ST mass function to avoid expensive solution of integral equations at many values of $m_{a}$. In future work, we will assess the impact of this approximation on our limits to ULA properties. Mass functions for a representative range of $m_{a}$ values are shown in Figure \ref{fig:duCutoffs}, demonstrating the suppression of the halo mass function (HMF) compared to CDM for $M_h\ll M_{J}(m_a)$. 

With dark-matter halo number densities in hand, we turn to the simulation of realistic HI surveys to evaluate their utility in testing ULA DM. For the simulations discussed below, we generated HMFs using $11$ logarithmically spaced values in the range $10^{-25}~{\rm eV}\lesssim m_{a}\lesssim 10^{-20}~{\rm eV}$, filling in the $m_{a}$ range near our ultimate lower limit with a denser grid of $20$ values satisfying $10^{-22.9}~{\rm eV}\lesssim m_{a}\lesssim 10^{-21.1}~{\rm eV}$.

\section{Simulation and analysis methods} \label{sec:methods} 

In this section we describe how we generate mock HI surveys in both simulated $\Lambda$CDM and axion dark matter universes. 

\subsection{Generating halos}

We populate spherical universes $100\,$Mpc in radius according to the HMFs of the CDM and axion models. Each HMF model, $\Phi$, is equidistantly binned in logarithmic mass space such that the expected number of halos in a given mass bin $i$ is
\begin{equation}\label{eq:n_per_bin}
    N_i = \Phi(m_i) (\Delta \log M) V,
\end{equation}
where $\Delta \log M$ is the logarithmic distance between mass bins and $V$ is the volume of the universe. Assuming the mass distribution of halos is Poissonian, we determine the number of halos per mass bin by sampling a Poisson distribution with an expected value determined by Equation \ref{eq:n_per_bin}.

For each halo, we select a random position within the sphere describing our mock universe and assign a randomly sampled disc inclination angle, $i$ (from a uniform $\sin i$ distribution). While the spatial distribution of galactic halos in these universes is random, and therefore unrealistically homogeneous, the halo mass distribution is an appropriate representation of the galactic population as predicted by CDM and ULA models.

\subsection{Assigning HI masses to halos} 

One of the most challenging steps in using HI surveys to constrain DM models is linking the HI detections to halo masses. While constraints on the baryon fraction of more massive halos ($M_h>10^{11}\, \mathrm{M}_\odot$) have been obtained from a variety of methods, complicated baryon physics, including stochastic events like supernova feedback, questions about the impact of processes during the epoch of reionization, and observational limitations, all complicate the expected baryon content of lower mass halos. Below $M_h\sim10^8\, \mathrm{M}_\odot$, any dark matter halos that exist are not expected to contain significant baryons \citep{Buckley2018}, so the halo mass range of particular interest is $M_h\sim10^{8-11}\, \mathrm{M}_\odot$ (probing ULA models with $m_a\sim10^{-(22-24)}\,$eV, see Figure \ref{fig:duCutoffs}).

\begin{figure}
    \includegraphics[width=\columnwidth]{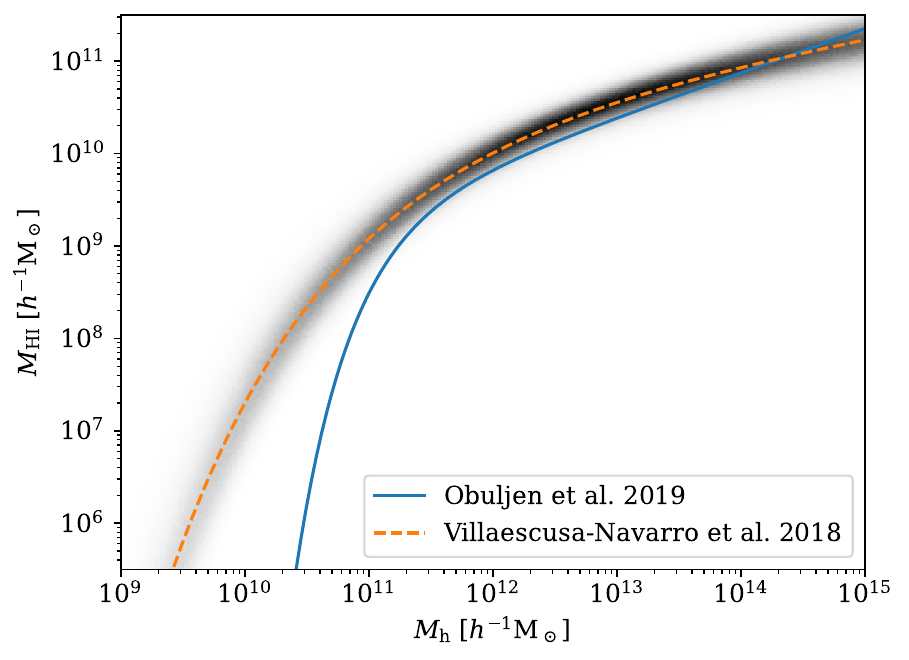}
    \caption{The $M_h$-to-$M_{\text{HI}}$ relationships predicted by \citet{Villaescusa-Navarro2018} (dashed orange line) based on the TNG100 simulation. We use this relation here to fill simulated halos with HI. The greyscale heat map shows simulated galaxy HI masses corresponding to $10^7$ uniformly sampled halo masses; each halo in our mock universe is assigned a corresponding HI mass assuming this model, with normally distributed uncertainties on fit parameters. Also shown is the relation measured by \citet{Obuljen2019} (solid blue line) based on ALFALFA.}
    \label{fig:MhMHIModels}
\end{figure}

In this work we use the $M_h-M_{\rm HI}$ function in \citet{Villaescusa-Navarro2018}, which is based on output from Illustris TNG (specifically TNG100; \citet{NelsonTNG}, a magnetohydrodynamic simulation of a $75\,h ^{-1}\,$Mpc box) from IllustrisTNG. They describe this using a functional form
\begin{equation} \label{eq:VNNew}
    M_{\text{HI}}(M_h) = M_0 \left(\frac{M_h}{{M_\text{min}}}\right)^\alpha \exp{\left(-\frac{M_{\text{min}}}{M_h}\right)^{0.35}},
\end{equation} where $M_0=(4.3\pm1.1)\times 10^{10} h^{-1} \mathrm{M}_\odot$, $\alpha=0.24\pm0.05$ and $M_{\rm min}=(2.0\pm0.6)\times 10^{12} h^{-1} \mathrm{M}_\odot$. 
For each mock universe, we fill halos of a given DM mass with HI according to this relation, selecting model parameters assuming they are normally distributed around the quoted values with the quoted uncertainties. An example result of this is shown in the greyscale heat map in Figure \ref{fig:MhMHIModels}, while the \citet{Villaescusa-Navarro2018} model is shown with the dashed orange line. For a discussion of how the choice of HI--halo mass conversion might impact our final result see Section \ref{sec:caveats}.

\subsection{Mock HI observations and HIMF construction}

To conduct mock observations, we determine the detectability of each halo's HI content given a set of survey sensitivity parameters. HI survey sensitivity depends both on total flux and line width, since the same amount of HI spread out over a wider velocity range will be harder to detect.

Typically the flat part of a galaxy rotation curve, as probed by the width of the HI global profile, is not at a radius which encloses all of the total halo mass. We therefore do not go directly from halo mass to rotation width using the laws of gravity, but instead assign the disc rotation velocity using the $v_{\rm flat}$--$M_h$ relation found by \citet{Katz2019}, namely
\begin{equation}
    V_{\text{flat}} = 10^{-B/A} M_h^{1/A}
\end{equation}
where they find $A=2.902\pm0.138$ and $B=5.439\pm0.292$ based on a fit to the SPARC dataset (a sample of 175 nearby galaxies with Spitzer data and resolved rotation curves \citealt{Lelli2016}), with halo profiles model as in \citet[][DC14]{DC14} which they found to provide a better fit than the traditional NFW profile. 

This $V_{\text{flat}}$ is then used to obtain the ``true'' HI line width of ${W_{\text{true}}=2V_{\text{flat}}}$. The total integrated HI flux $S_{21}$ is largely invariant over disc inclination angle, $i$, since at most radii HI is optically thin, so there is minimal impact from self-absorption at 21cm (e.g. \citet{Springob2005} find self-absorption corrections to be less than 30\% even in the most inclined systems), however observed line widths are strongly dependent on inclination as only the line-of-sight velocity generates a Doppler shift, such that $W_{\text{obs}} = W_{\text{true}} \sin{i}$.

 We use analytic approximations of survey sensitivity for both the completed Arecibo Legacy Fast Arecibo L-band Feed Array survey (ALFALFA; \citealt{Haynes2018}), which used the Arecibo radio telescope to survey the entire extragalactic sky accessible from Arecibo in the 21cm line at the 3\arcmin ~resolution provided by Arecibo, and the ongoing Widefield ASKAP L-band Legacy All-sky Blind surveY (WALLABY; \citealt{Koribalski2020,Westmeier2022}) which is currently running on the Australian Square Kilometer Arrary Pathfinder (ASKAP) with a goal of surveying 75\% of the sky (the part accessible from Western Australia) in the 21cm line at 3\arcsec~ resolution.

For ALFALFA, \citet{Haynes2011} cite a 90\% completeness limit of
\begin{equation}
  \log{S_{21,90\%,\text{Code 1}}}=\begin{cases}
    0.5\log{W_{50}}-1.14, & \log{W_{50}}<2.5.\\
    \log{W_{50}}-2.39, & \log{W_{50}}\geq2.5,
  \end{cases}
\end{equation}
for their ``Code 1'' (reliable, signal-to-noise, $S/N>6.5$) detections, where the total HI flux, $S_{21}$ in Jy, is linked to HI mass (in $\mathrm{M}_\odot$) and the distance of the galaxy, $D$ (in Mpc) using the standard relation of,
\begin{equation}
    M_{\text{HI}} = \left(2.356\times10^5\right)D^2 S_{21}.
\end{equation}

For WALLABY, \citet{Koribalski2020} provided an estimated 5$\sigma $ $M_{\text{HI}}$ detection limit for a source measured by a single beam as a function of line width, $W$\footnote{It is not specified which measure of $W$ they assume.} in km s$^{-1}$ and distance $D$ in Mpc as
\begin{equation}
    M_{\text{HI}}\, [\mathrm{M}_{\odot}] = \left(3901W^{0.493}\right)D^2,
\end{equation}
which corresponds to $M_{\text{HI}}=5.3\times10^8\, \mathrm{M}_{\odot}$ at a fiducial distance of $100\,$Mpc and assuming typical $W=250\,$km s$^{-1}$. The pilot data release for WALLABY \citep{Westmeier2022} recently released measured sensitivity (not quoting any width dependence) of $M_{\text{HI}}=5.2\times10^8\, \mathrm{M}_{\odot}$ at $100\,$Mpc, which is very comparable. In what follows we continue to use the \citet{Koribalski2020} estimate. 

The effect of non-detections due to line-of-sight halo occultations should be negligible since, with HI, one can detect two halos along the same line of sight (LOS) as long as they are separated enough in redshift. If a mock galaxy with a width $W_{50}$ (hereafter just $W$) and HI mass $M_{\text{HI}}$ has a flux greater than these limits, we consider it detected in the relevant mock survey.

The output of this process is a catalogue of HI detections with halo masses, rotation widths, and HI masses. We make use of this to construct mock HI Mass Functions (HIMFs) using a $1/V_{\rm max}$ weighting methods, where $V_{\rm max}$ is the maximum volume in which a detection of a given HI mass could be observed. No corrections are applied for the impact of large scale structure (LSS) or distance errors on the HIMF since our mock universes are homogeneous and we have exact distances (we refer the reader to \citealt{Masters2004} for an explanation of these effects in real surveys). 

\subsection{Statistics of detectable differences from CDM}

In order to quantify how distinguishable our simulated measured HIMFs from mock ULA dark matter universes are from fiducial $\Lambda$CDM universes, we make use of the two-sample Kolmogoroff-Smirnoff (KS) test. This is a standard statistical technique for testing the likelihood that two distributions are samples of a single underlying population (i.e. that they are the same). We implement this using the SciPy Stats KS two-sample test package.\footnote{\url{https://docs.scipy.org/doc/scipy/reference/generated/scipy.stats.ks_2samp.html}}

\section{Results and Discussion} \label{sec:results}

\subsection{HIMF comparisons}
\begin{figure*}
    \includegraphics[width=0.8\textwidth]{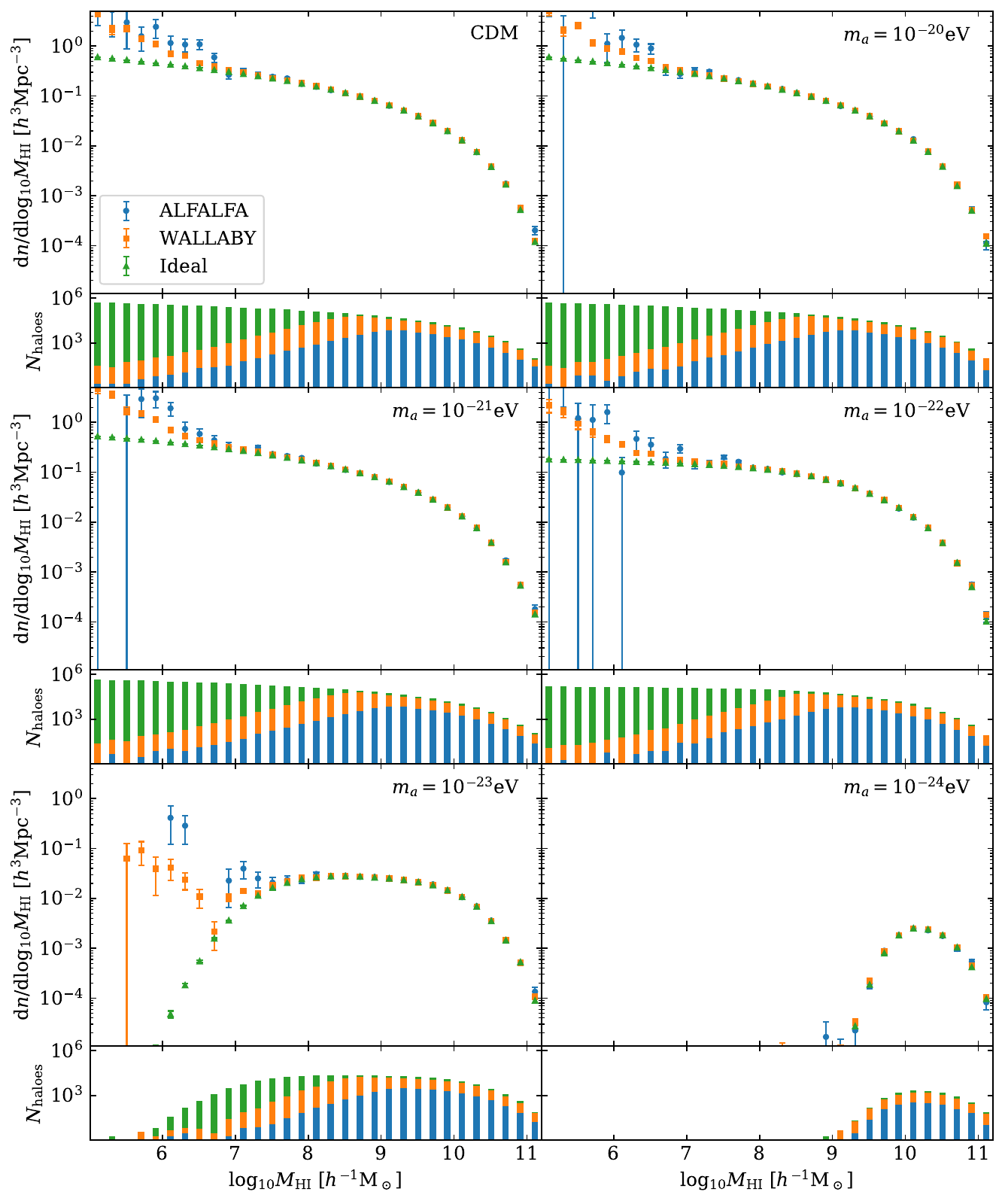}
    \caption{Mock HIMFs for a $100\,$Mpc survey in a Universe with dark matter that is either CDM or a ULA with $m_a=10^{-20},\ 10^{-21},\ 10^{-22},\ 10^{-23},$ and $10^{-24}\,$eV (as labeled). In each of the six subplots, we show in the upper panels the observed mock ALFALFA and WALLABY mass functions (blue circles and orange squares respectively), while green triangles correspond to the same from a hypothetical ``ideal'' survey able to observe all halos in the simulated universe (i.e. these points reflect the underlying true HIMF). Error bars account for Poisson counting error. The histograms in the lower panels show the total number of observed halos per mass bin in each survey (with the same colour scheme).}
    \label{fig:HIMF_multi}
\end{figure*}

We show the output HIMFs from five of our mock universes in Figure \ref{fig:HIMF_multi} corresponding to universes with dark matter comprised of a ULA with a mass $m_a=10^{-24},\ 10^{-23},\ 10^{-22},\ 10^{-21},\ 10^{-20}\,$eV and a universe with standard CDM. Three versions of the HIMF (as both a histogram of raw number counts in the lower part of each panel, and corrected for volume in the upper part) are show for each mock universe. We show an HIMF from an ideal survey (which detects all halos with HI mass as per the $M_{\rm HI}$--$M_h$ relation (blue), as well as HIMFs generated from our mock ALFALFA and WALLABY surveys in each mock universe (blue and orange respectively). 

The three measurements of each universe agree within the error bars at high masses, but deviations are seen at low mass, caused by a Malmquist bias \citep{Malmquist1922,Malmquist1925} type effect. Since at these masses, the majority of galaxies are below or close to the sensitivity limit of the surveys, halos with more HI than the mean trend are more likely to be detected, and since the HIMF rises (in log-space) towards lower mass halos this results in a significant turn up in the measured HIMF. This observation shows the importance of comparing mock observations of simulations to real measurements in making such comparisons (i.e. forward modelling). 

\begin{figure*}
    \includegraphics[width=\columnwidth]{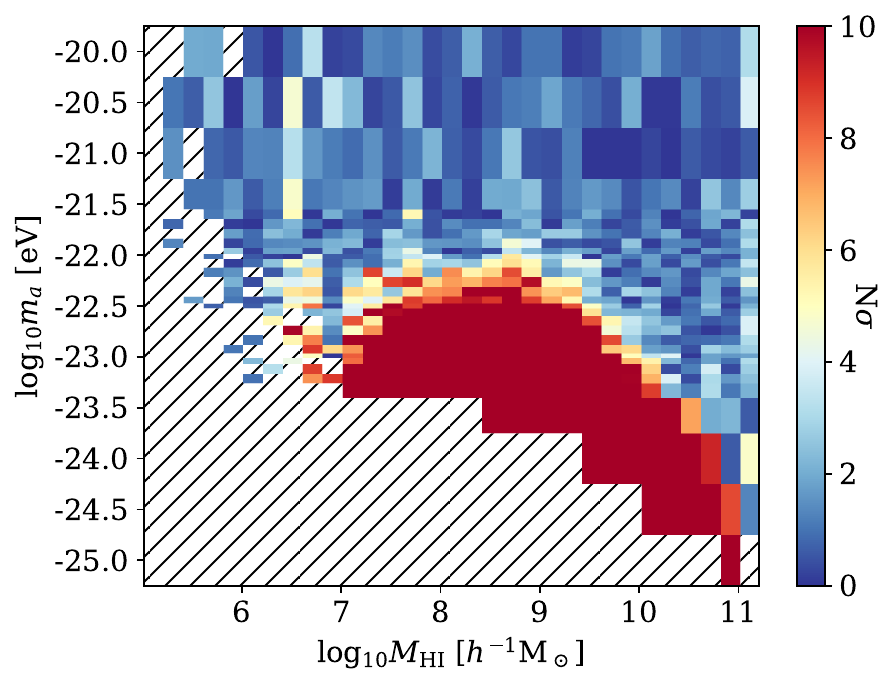}
    \includegraphics[width=\columnwidth]{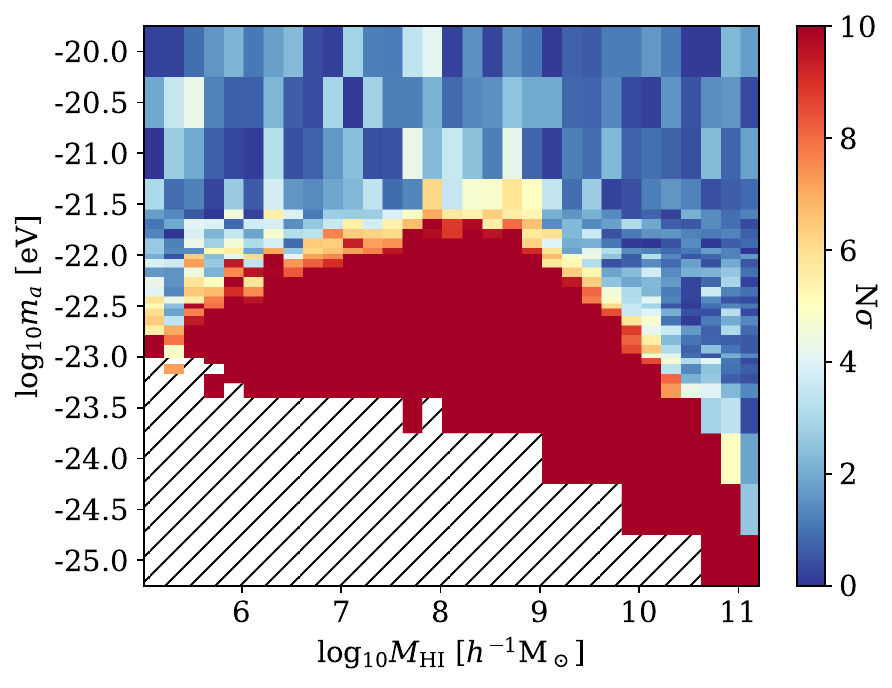}
    \caption{Deviation between ALFALFA (left) and WALLABY (right) mock HIMFs and the observed CDM HIMF in terms of observational uncertainty $\sigma$. Regions of parameter space with three or fewer HI detections in the mock surveys are shown by the slashed regions.}
    \label{fig:deviation_ALFALFA}
\end{figure*}

We quantify the detectability of a given ULA cosmological model by considering the amount by which the observed HIMF can be seen to deviate from CDM. This deviation is calculated per mass bin in terms of $\text{N}\sigma$ as $(\Phi_\text{obs} - \Phi_\text{CDM}) / \sigma_\text{obs}$ where $\Phi_\text{obs}$ and $\sigma_\text{obs}$ are the observed HIMF and corresponding uncertainty and $\Phi_\text{CDM}$ is the observed CDM HIMF. The deviations from CDM as observed by our mock ALFALFA and WALLABY surveys are shown in Figure \ref{fig:deviation_ALFALFA}.

These detectable deviations from CDM in our mock universes suggest that both ALFALFA and WALLABY can constrain models with ULA masses below $10^{-22}\,$eV, while WALLABY has the potential to constrain ULAs in the mass range up to $10^{-21}\,$eV. 

\begin{figure}
    \includegraphics[width=\columnwidth]{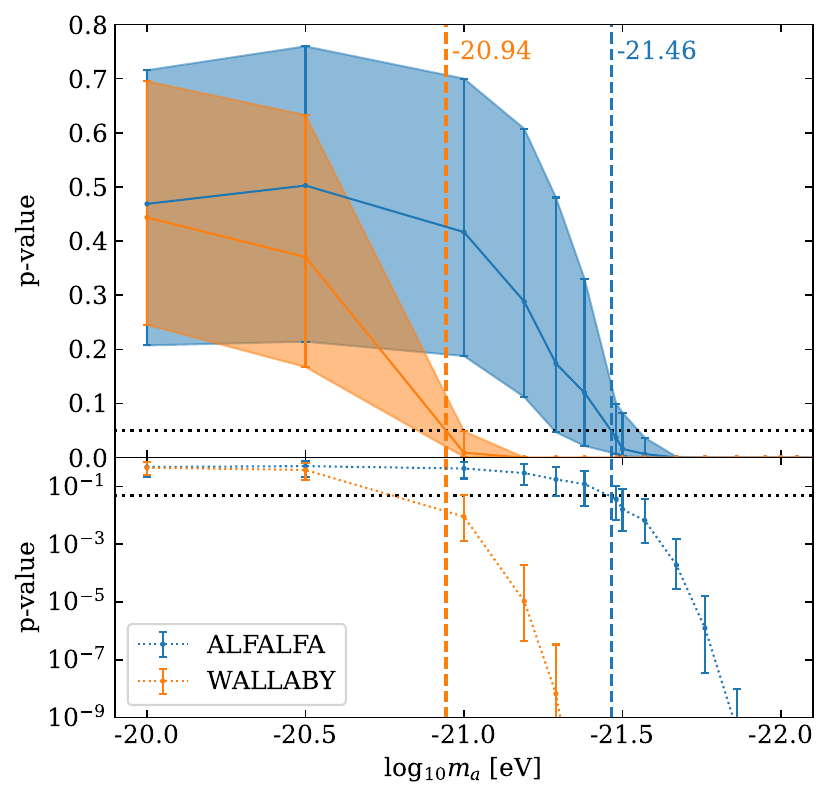}
    \caption{Linear (top) and logarithmic (bottom) display of the median and inter-quartile range (IQR) of KS-test p-values comparing mock ALFALFA (blue) and mock WALLABY (orange) surveys in ULA dark matter universes to surveys in mock $\Lambda$CDM universes. This shows the distribution of 100 sets of simulated universes in both $\Lambda$CDM and each of the axion dark matter masses. The horizontal dotted line indicates a p-value of 0.05, the typical threshold used to indicate statistically significant different distributions. Vertical dashed lines mark the inferred masses at which the surveys cross the threshold.}
    \label{fig:KSvalues}
\end{figure}

 Another way to test for the deviation of axion dark matter models from CDM is via a KS test. We show this in Figure \ref{fig:KSvalues}, which plots the median and inter-quartile ranges of $p$-values from a series of independent KS tests comparing the HI halo mass distributions from mock ULA universes (with particle mass as shown on the $x$-axis) with those from our mock $\Lambda$CDM universes in both mock ALFALFA and mock WALLABY surveys. A $p$-value$<0.05$ (indicated by the horizontal dashed line) is typically considered to show statistically different distributions. We find that (averaged over 100 simulations) the KS test suggests statistically different HI halo mass distributions between ULA universes compared to $\Lambda$CDM for ULA masses $\log_{10}{\left(m_a/{\rm eV}\right)} > -20.9\pm0.2$ for the mock WALLABY survey and $\log_{10}{\left(m_a/{\rm eV}\right)} > -21.5\pm0.2$ for the mock ALFALFA survey.

\subsection{Caveats and limitations}\label{sec:caveats}
There are two ways in which our predicted constraints on the ULA mass from these mock HIMF reconstructions may be optimistic:
\begin{enumerate}
    \item We have limited our study to cosmologies in which dark matter is exclusively composed of axions. However, some mixed dark matter models that take axions to be a fraction of all dark matter display MSP-resolving halo cutoffs as well \citep{Marsh2014}. Despite differences in other halo properties, alternative models may be subjectable to the analysis considered in this paper if they demonstrate similar HMF cutoff behavior.
    \item The adopted $M_{\rm HI}$--$M_h$ relation is based on simulated data and appears to over-represent the HI content of low-mass galaxies. There are a variety of techniques to estimate and/or measure the HI--halo mass function. While we base this work on a predicted HI--halo mass function based on a simulation \citep{Villaescusa-Navarro2018}, it is also possible to estimate the function from real data (with significant error). For example, using ALFALFA data, \citet{Obuljen2019} fit the Schechter function, 
\begin{equation}
    M_{\text{HI}}(M_h) = M_0 \left(\frac{M_h}{{M_\text{min}}}\right)^\alpha \exp{\left(-\frac{M_{\text{min}}}{M_h}\right)}, 
\end{equation}
with values of $\log_{10}{(M_0/h^{-1}\mathrm{M}_\odot)}= 9.44^{+0.31}_{-0.39}$, $\log_{10}( M_{\text{min}}/h^{-1}\mathrm{M}_\odot)=11.18^{+0.28}_{-0.35}$, and $\alpha=0.48\pm0.08$. We show a comparison of this with the model we use in this work in Figure \ref{fig:MhMHIModels}. The two lines show good agreement at higher masses, diverging from each other on low-mass scales (below about $M_h\sim10^{11}\, \mathrm{M}_\odot$, or $M_{\rm HI}\sim10^9\, \mathrm{M}_\odot$). 
    An overprediction of the amount of HI in a halo mass of a given size seems to be fairly common in simulations. This discrepancy is not unexpected between observations and a simulation, since modeling gas physics, and gas phases is particularly challenging. For example, IllustrisTNG may not account for all sources of ionizing radiation, which would impact the amount of HI able to survive in lower-mass halos. \citet{LiX2022} look at this question in several recent simulations. \citet{Thananusak2023} consider the impact this uncertainty has on the match between $\Lambda$CDM and simulated HIMFs. While our mock universes are self-consistent, our choice of $M_{\rm HI}$--$M_h$ likely means our mock surveys contain a larger number of HI detections than the real ones can expect, resulting in underestimated Poisson counting errors.  
    
    While the HI--halo mass relation is often presented as a mean trend, it is well known that halos of the same mass can host a variety of HI abundances (and that this has interesting astrophysical reasons). For example, \citet{Guo2020} used a stacking analysis to directly measure the mean HI mass in galaxy groups with a range of halo masses, demonstrating this wide variation. Understanding the details of the HI bias (e.g. as in \citealt{Guo2023}) in filling halos is interesting both for astrophysics and the use of HI as a cosmological probe (e.g. HI intensity mapping), but beyond the scope of this paper. 
\end{enumerate}

\subsection{Comparison with real data}

\begin{figure}
    \includegraphics[width=\columnwidth]{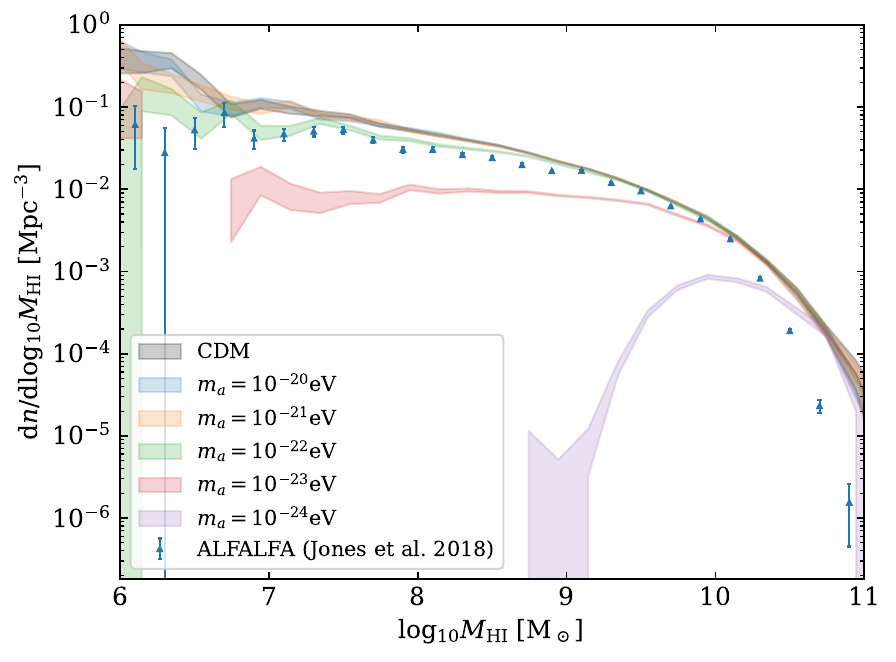}
    \caption{A set of mock ALFALFA HIMFs from universes with different models for dark matter, compared to the real ALFALFA HIMF. The mock HIMFs as observed by ALFALFA are shaded within Poissonian error range. The blue triangles correspond to the ALFALFA HIMF from \citet{Jones2018}. The mock HIMFs have been corrected to be in terms of $h_{70} = 1.0$ to allow for physically meaningful comparison with the real ALFALFA data. Based on this, there appears to be reasonable agreement with observed data in ALFALFA and CDM, as well as all ULA models with $m_a>10^{-22}~{\rm eV}$. ULA models with $m_a<10^{-23}~{\rm eV}$ clearly do not match the data.}
    \label{fig:real_ALFALFA_comparison_ALFALFA}
\end{figure}

Since the ALFALFA survey is already complete \citep{Haynes2018}, we can also test the ULA model against a real measured HIMF from ALFALFA. We use the HIMF published in \citet{Jones2018} and compare it to our mock ALFALFA surveys from CDM and a range of ULA universes in Figure \ref{fig:real_ALFALFA_comparison_ALFALFA}. While the agreement between the real ALFALFA HIMF and our mock universes is not perfect, particularly at the highest masses, we note that this is quite likely due to uncertainty in the $M_{\rm HI}$--$M_h$ relation we adopt. Indeed, the prescription of \cite{Villaescusa-Navarro2018} is known to not match the high-mass HMF in observations, and a definitive observational limit would require some modified prescription to mitigate this mismatch \citep{Guo2020,2022arXiv220710414L}, accounting for any number of baryonic effects (e.g. ram-pressure stripping). 

Around the knee and towards lower masses, the shape of the HIMF from ALFALFA already appears to rule out ULA (as the sole DM particle) if they have masses $m_a<10^{-23}\,$eV, however we note that dark matter comprised of ULA masses with $m_a=10^{-22}\,$eV or larger are nearly indistinguishable from CDM in this test. All of our model universes with ULA masses larger than this as well as CDM agree reasonably well with the observations from ALFALFA. 

\section{Summary and Conclusions} \label{sec:summary}

The research described in this {work} was motivated by a wish to explore ULA models for dark matter. This type of axion dark matter suppresses small scale structure, and might thus help to explain some of the well known inconsistencies between standard $\Lambda$CDM and observations on low mass scales.

We develop a method by which to test the potential for low-redshift large area HI surveys to constrain ULA models for dark matter via the shape of the HIMF. This involves making mock universes populated with halos from different HMFs, filling them with HI using a $M_{\rm HI}$--$M_h$ relation, and then ``observing'' them to the sensitivity levels of recent/upcoming surveys. 

We generate mock HIMFs based on universes filled with dark matter modeled as several different ULA particle masses and also CDM. We do this for both an ideal survey (i.e. detecting everything), and mock surveys designed to mimic the sensitivity of both the completed ALFALFA \citep{Haynes2018} and ongoing WALLABY surveys. \citet{Westmeier2022} recently released WALLABY pilot data demonstrating the survey was able to reach the nominal sensitivity limits presented in \citet{Koribalski2020} (and which we used in our simulations).

We find that the HIMF measured from ALFALFA can clearly be used to detect ULAs with $m_{a}\leq 10^{-21.5}~{\rm eV}$, while the future WALLABY survey will reach the window
$m_{a}\leq 10^{-20.9}~{\rm eV}$. This work demonstrates the promise of HIMFs measured from upcoming and/or existing large area local Universe HI surveys to place constraints on ULA models for dark matter in the range of $m_{a}\sim 10^{-22}-10^{-21}~{\rm eV}$. 

\begin{figure}
    \includegraphics[width=\columnwidth]{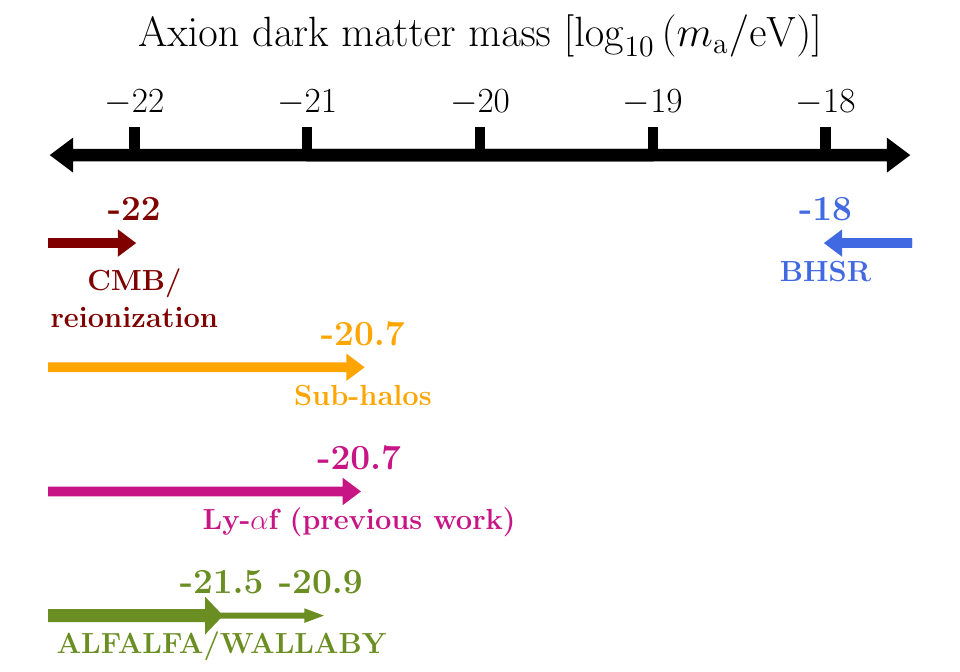}
    \caption{Mass ranges tested by a variety of astrophysical probes, including those discussed in this work.}    \label{fig:mass_domain}
\end{figure}

 To place our constraints in context, we show the ULA mass range probed by a variety of other astrophysical constraints in Fig.~\ref{fig:mass_domain}. The most robust constraints come from CMB anisotropies (including lensing and its impact on polarization) and the optical depth to reionization \citep{2019BAAS...51c.567G}. Along with black hole spin population statistics (marked by the black hole super radiance or BHSR exclusion region), this highlights a region of interest with $10^{-22}~{\rm eV}\lesssim m_{a}\lesssim 10^{-18}~{\rm eV}$ \citep{2019BAAS...51c.567G}. Additional (promising, but subject to considerable astrophysical uncertainty) constraints include the sub-halo mass function (as probed by strong gravitational lensing), Lyman-$\alpha$ forest absorption of QSO emission, and then our own work (see \citealt{2019BAAS...51c.567G,2020PhRvD.101l3026S,2021PhRvL.126g1302R} and references therein). Our work complements considerations of using the spatial statistics of HI line-intensity mapping to test for ULA DM \citep{Hotinli:2021vxg,2021MNRAS.500.3162B,2021PhRvD.103b3521K,2021ApJ...913....7J,2021IJMPD..3030009P,2022JCAP...08..066K,2022PhRvD.106f3504F}. HI data have also been applied to tests of high-mass axions in the range, $10^{3}~{\rm eV}\lesssim m_{a}\lesssim 10^{5}~{\rm eV}$, by considering the impact on gas cooling rates in the Leo T dwarf galaxy \citep{WadekarWang2022}.

Since the ALFALFA survey is complete, we are additionally able to compare our model to the real HIMF \citep[we use the one published in][]{Jones2018}. This demonstrates reasonably good agreement with our mock CDM model, as well as those with ULA masses $m_a=10^{-22}\,$eV and larger, while appearing to rule out ULA dark matter with masses $m_a=10^{-23}\,$eV and smaller. 

One strong motivation for considering ULAs in the mass range we explore is the \emph{axiverse}, which posits many ultra-light fields inspired by string compactifications. As theoretical tools to model linear and non-linear structure formation in these scenarios become available (e.g. \citealt{Chen:2023unc} and \citealt{Luu:2023dmi}), we look forward to exploring the impact on and constraining power of HI surveys to test these possibilities.

\section*{Acknowledgements}
The authors acknowledge support from the Provost's office at Haverford College through Summer Student Stipends. D.~G. thanks K.~K.~Rogers for providing the original script that was modified to produce Fig.~7. D.~G. acknowledges support in part by NASA ATP Grant No. 17-ATP17-0162 and the Kaufman New Initiative research grant KA2022-129518. We acknowledge useful conversations with D.~J.~E.~Marsh, X.~Du, M.~Haynes, A.~Lidz, J.~Sipple, and R.~Ghara.

The Arecibo Observatory is operated by SRI International under a cooperative agreement with the National Science Foundation (AST-1100968), and in alliance with Ana G. Méndez-Universidad Metropolitana, and the Universities Space Research Association.

The land on which the Haverford College stands is part of the ancient homeland and unceded traditional territory of the Lenape people. We pay respect to Lenape peoples, past, present, and future and their continuing presence in the homeland and throughout the Lenape diaspora. 

\section*{Data Availability}


Simulated data available on request to the authors.



\bibliographystyle{mnras}
\bibliography{main}








\bsp	
\label{lastpage}
\end{document}